\documentclass[12pt]{iopart}
\usepackage[english]{babel}

\expandafter\let\csname equation*\endcsname\relax
\expandafter\let\csname endequation*\endcsname\relax

\usepackage{graphicx}
\usepackage{bm}
\usepackage{physics}
\usepackage{subfigure}
\usepackage{color}
\usepackage{hyperref}

\newcommand{\red}[1]{#1}

\begin{document}

\title{The effect of substrate waviness on random sequential adsorption packing properties}

\author{Piotr Kubala, Micha\l{} Cie\'sla}
\address{Institute of Theoretical Physics, Department of Statistical Physics, Jagiellonian University, \L{}ojasiewicza 11, 30-348 Krak\'ow, Poland}

\date{\today}

\begin{abstract}
\red{Random sequential adsorption of spheres on a wavy surface was studied. It was} determined how surface structure influences random packing properties such as the packing fraction, the kinetics of packing growth and the two-particle density correlation function. Until the substrate varies \red{within the range} one order of magnitude smaller than the particle's diameter, the properties of the packings obtained do not differ significantly from those on a flat surface. On the other hand, for the higher amplitude of unevenness, the packing fraction, low-density growth kinetics and the density autocorrelation function change significantly, while asymptotic growth kinetics seems to be barely sensitive to surface waviness. Besides fundamental significance, the study suggests that the experimental measurement of the aforementioned basic properties of adsorption monolayers can reveal the surface's porous structure without investigating the surface itself. 
\end{abstract}

\maketitle

\section{Introduction}

The study of adsorption and deposition of various molecules \red{on to} solid or liquid interfaces is of major significance for both fundamental and applied sciences. \red{It contributes to the first by allowing to determine} adsorbate physical properties, such as its shape or charge distribution, and \red{explore in detail the course} of interactions binding a molecule to \red{an} interface. On the other hand, \red{certain} specific properties of adsorption monolayers \red{may be applied directly} in material, food and medical sciences \red{as well as in} pharmaceutical and cosmetic industries. For example, adsorption phenomena are involved in blood coagulation, artificial organ failure, plaque formation, inflammatory response, fouling of contact lenses, ultrafiltration, and are \red{utilised} in membrane filtration units. Controlled deposition on various surfaces is a prerequisite of efficient separation and purification by chromatography, gel electrophoresis, filtration, biosensing, bioreactors, immunological assays, etc. \cite{Adamczyk2006, Dabrowski2001, Adamczyk2012}. 

\red{Numerous experiments focus} on the adsorption on flat surfaces e.g., mica \cite{Wasilewska2011, Fitzpatrick1992}. However, there are many systems where deposition occurs on patterned, rough, or curved surfaces \cite{Pfeifer1989, Li2002, Herminghaus2012,Memet2019}. Results of these studies show that the character of an interface can significantly modify adsorption layers properties. Likewise, in numerical modelling of such processes, new methods should be used to simulate molecules deposition effectively \cite{Chen2017}.

One of the simplest and most popular \red{protocols} used for modelling monolayers \red{created} during irreversible adsorption experiments is random sequential adsorption (RSA) \cite{Feder1980, Evans1993}. The algorithm \red{is based} on repeated random sampling of the adsorbate molecule position and orientation on \red{a} surface or \red{an} interface, followed by the check if the molecule does not intersect with other molecules that are already placed in the neighbourhood. If there is no intersection, the molecule is \red{kept} in this position and otherwise it is removed.
\red{As} RSA is \red{a} very rough approach to modelling adsorption monolayers, it \red{led to} various generalisations. The most popular one is ballistic deposition \cite{Talbot1992, Choi1993, Schaaf2000}. If a trial particle hits an already placed particle, it slides down to the surface and stays there. This model is \red{particularly} useful when deposited particles are relatively heavy and gravity dominates their movement. The packing fractions obtained in the ballistic RSA are typically $5-10\%$ higher than \red{the} ones obtained in the standard RSA \cite{Schaaf2000}. Another approach proposed by Senger et al. \cite{Senger1992} assumes the diffusion to be the crucial transport mechanism and simulates the Brownian motion of depositing molecules. It appears that in this model, the kinetics of packing growth changes, but the obtained packing fractions are the same as in the standard version of RSA. The common drawback of all these approaches is their inefficiency when the obtained monolayers are nearly saturated -- there is little space left for subsequent additions. \red{All of the above algorithms require then} a large number of trials to find a suitable place for a single particle.

RSA is also known as one of the simplest yet not trivial protocol for generating random packings taking into account excluded volume effects \cite{Torquato2010}. For this reason, it is important and interesting from the perspective of random packings and properties of granular and random media in general. Therefore \red{many} of theoretical and numerical studies were performed to \red{investigate} the properties of RSA packings of various objects \cite{Evans1993}.  

\red{This manuscript presents the results of numerical simulations of} RSA monolayers/packings built of spheres placed randomly on a wavy surface. \red{For the purposes of the study, an algorithm was developed that} allows to generate such saturated RSA packings in \red{a} finite simulation time \red{with} a guarantee that no additional sphere can be placed \red{on the investigated surface} without changing the positions of already deposited objects. The properties of the obtained monolayers are analysed in terms of packing fraction, density autocorrelation function, and packing growth kinetics. \red{It is especially interesting} how the period and the amplitude of surface folding influence these properties. One of this study's goals is to find \red{whether} it is possible to determine the surface structure by \red{analysing solely} the essential characteristics of the modelled monolayers. A similar analysis has been performed for surfaces \red{resembling} meshes \cite{Ciesla2017}.

\section{Model}

\begin{figure}
    \centering
    \includegraphics[width=0.7\linewidth]{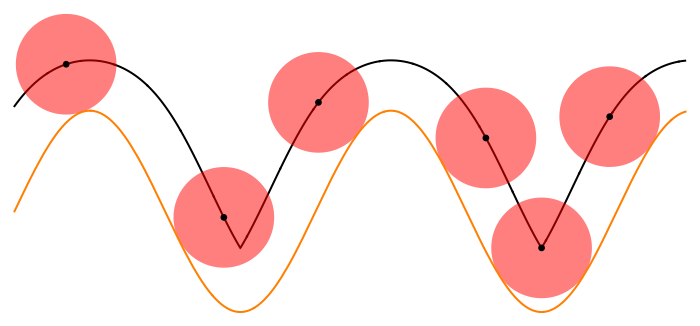}
    \caption{An illustration of the concept of a virtual surface. The actual surface is depicted as an orange (lighter) line, while a virtual as a black (darker) line. If centres of particles lie on a virtual surface, the particles are tangent to the real one.}
    \label{fig:surface}
\end{figure}

\red{The} heterogeneous substrate model \red{was} a sinusoidal surface of a linear size $L \approx 80$ oscillating in $x$ direction and constant in $y$ direction, defined by the equation
\begin{equation}
    z(x, y) = A \sin \qty(\frac{2\pi x}{\lambda}),
\end{equation}
for $x, y \in [0, L)$. Parameter $A$ \red{controls} the transverse amplitude of waviness, while $\lambda$ -- \red{its} longitudinal size. Adsorbate balls were of a unit diameter \red{$D = 1$}. \red{Thus, within the scope of interest of this study were the ranges of parameters} \red{$\lambda < D = 1$ and/or $A < D/2 = 1/2$, for which the} surface inhomogeneities are smaller than \red{the} adsorbed particles.

Monolayers of balls were created using random sequential adsorption (RSA) protocol \cite{Feder1980, Evans1993}. A single iteration of the RSA algorithm \red{proceeds} as follows:
\begin{itemize}
    \item select a random position of a ball uniformly in $xy$ plane;
    \item find numerically $z$ position of the ball centre so that it is tangent to the surface;
    \item check if it overlaps any previously adsorbed particle. If it does not, permanently add the ball to the packing; discard otherwise.
\end{itemize}

Note that the balls' centres lie on the surface given by the set of points distant from the real surface by the ball's radius. This surface will be called \textit{a virtual surface} in the \red{further} part of the manuscript (see Fig.\ref{fig:surface}). The main parameter tracked during simulation is the mean packing fraction (packing density) $\theta(N)$ depending on the instantaneous number of iterations $N$. The process is iterated until there is no space large enough to adsorb another particle. Such \red{a} packing is called saturated. The mean saturated packing fraction \red{was investigated more closely} in RSA studies \cite{Wang1994, Vigil1989, Brosilow1991, Ricci1992, Zhang2013, Ciesla2016}. Formally it is equal to $\theta(\infty)$. From now on, when $N=\infty$ \red{is referred to,} the argument $\theta\equiv\theta(\infty)$ \red{is dropped}. In the case of a wavy surface, the packing fraction \red{may} be defined in multiple ways. \red{The investigators} have chosen to work with two-dimensional projection, so
\begin{equation}\label{eq:theta}
    \theta = k S_P/S_\text{flat},
\end{equation}
where $S_P = \pi$ is the surface of projection of a ball on $xy$ plane, $S_\text{flat} = L^2$ is the area of surface assuming it is flat and $k$ is the mean number of shapes in the packing. Note that for \red{even} non-overlapping balls, their projections may overlap; \red{thus,} the packing fraction defined \red{in the above way} may exceed $1$ for a high local surface slope.

\section{Numerical simulations}

To generate saturated packings, a slightly modified method \red{was used} of tracing available regions where particles can be placed, described in \cite{Wang1994, Ebeida2012, Zhang2013} and \red{the authors'} previous studies \cite{Haiduk2018, Kasperek2018}. \red{In the summary, the  three-dimensional packing} is initially divided into identical cubic voxels. All voxels \red{entirely} contained in one of the particles' exclusion zone (a ball of two times larger radius) are marked inactive. New particles are sampled only from active voxels, not entirely contained in \red{the} exclusion zone. When the number of unsuccessful adsorption trials exceeds \red{certain} threshold value each voxel is divided into eight smaller ones. Then, each of them is rechecked if it \red{is} active. This increases the efficiency of sampling and enables \red{assessing} packing saturation -- if the packing is saturated, all voxels \red{are} marked inactive. This method \red{needed adaptation to the investigated} model -- particles' centres could be adsorbed only on the virtual surface, so all voxels that did not intersect the virtual surface were immediately marked inactive. Thus, the voxels were eventually discarded either because they lay outside the surface or were covered by the exclusion zone of a sphere that \red{had} already \red{been} placed in the packing.

The properties of packings \red{were investigated} for various pairs of parameters $A$ and $\lambda$ for $A \in [0.008, 8]$ and $\lambda \in [0.08, 80]$, notably \red{spreading} both smaller and larger length scales compared to particle diameter $D$. As \red{the paper will present further on, reasonably large deviations were noted, compared to} the flat-surface case, in some packing characteristics for surface waviness on up to one order smaller scale than the ball size. For each set of parameters, $10^3$ independent packings were generated for better statistics and to estimate the statistical error. The standard deviation was estimated assuming \red{that} packing fractions in individual packings follow a normal distribution, \red{which is} justified by previous studies \cite{Pedersen1993, Ciesla2017scaling, Ramirez2019, Baram2021}.

To minimise finite-size effects, periodic boundary conditions were used \cite{Ciesla2018}. Notably, they had to be compatible with the surface period, forcing $\lambda = L/n$, where $n$ is a positive natural number. For a flat surface case, \cite{Ciesla2018} indicates that for $D=1$, the substrate size $L=5$ is enough to keep finite-size error of $\theta$ below $10^{-5}$, thus $L=80$ used in the study gives an extensive margin. A wavy surface introduces long-range order; however, it does not originate from the excluded volume, so no additional finite-size effects are expected in the regime where $\lambda \approx L$. \red{It has been} confirmed by repeating the selected simulation for five times smaller surface with $L=16$ -- the results were identical within statistical error. For large $\lambda$ values, however, sampling density is restricted due to the condition $\lambda = L/n$. To \red{account for that, larger surfaces need to be considered}.

\section{Results and discussion}

\begin{figure}
    \centering
    \subfigure[]{\includegraphics[width=0.4\linewidth]{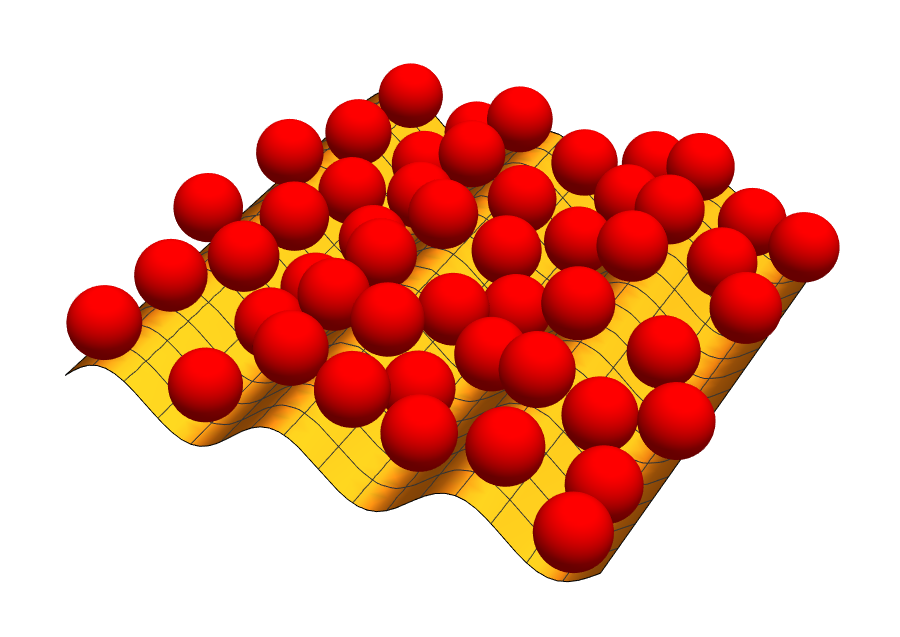}}
    \subfigure[]{\includegraphics[width=0.4\linewidth]{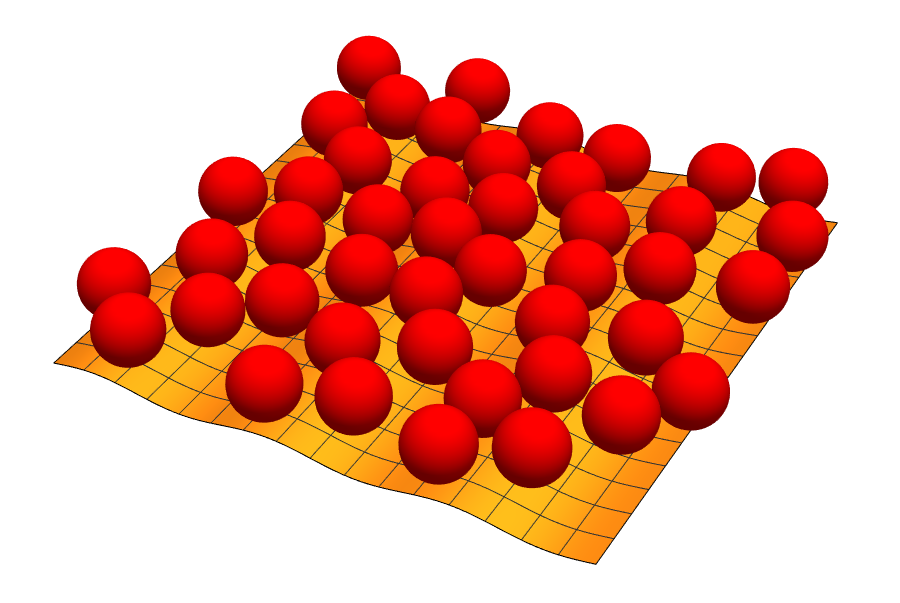}}
    \subfigure[]{\includegraphics[width=0.4\linewidth]{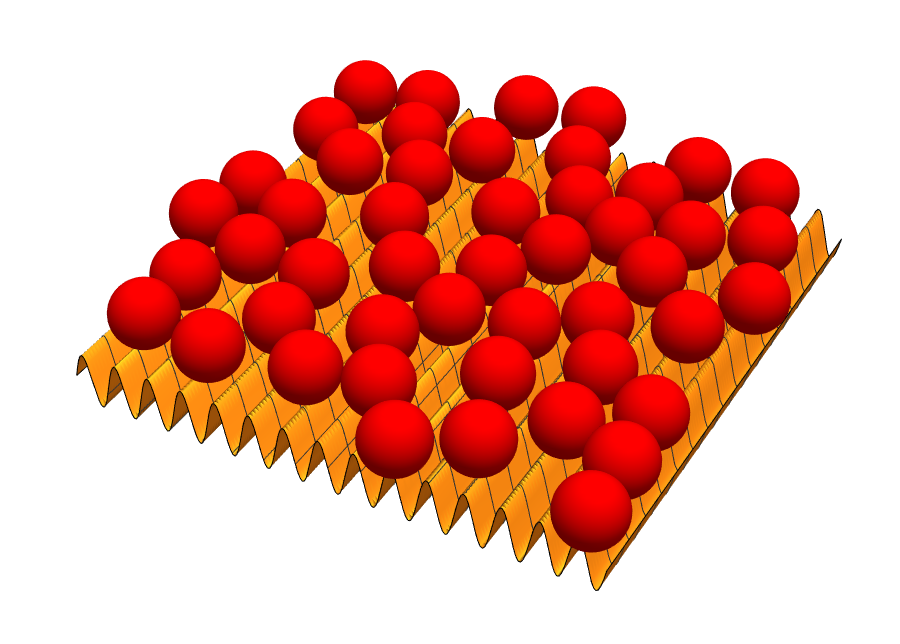}}
    \subfigure[]{\includegraphics[width=0.4\linewidth]{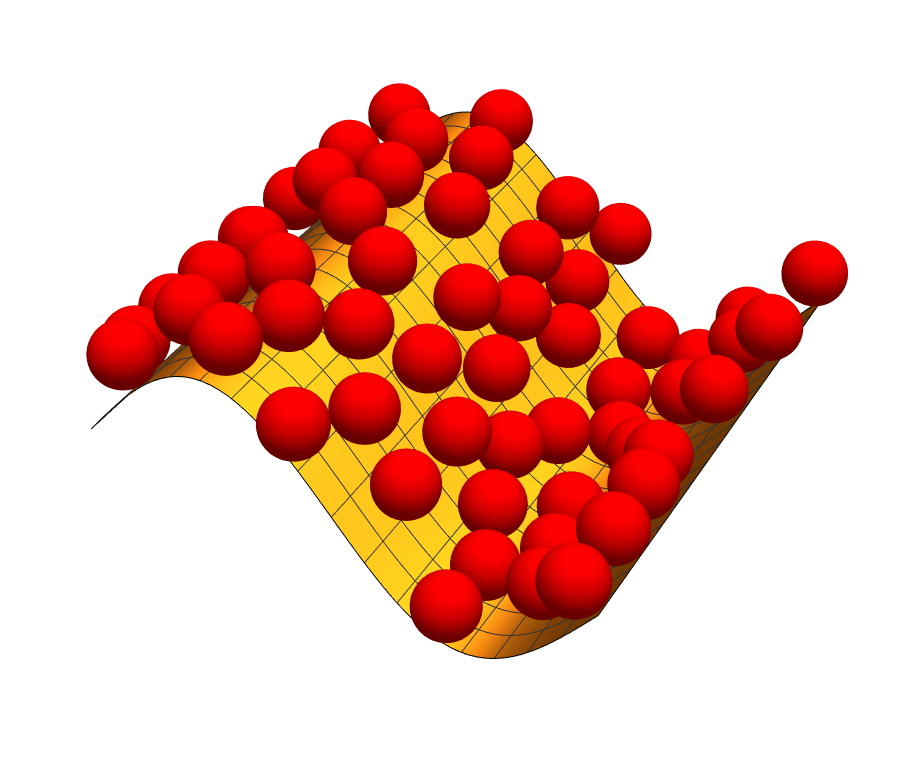}}
    \caption{Examples of saturated packings: (a) $(A, \lambda) = (0.4, 2.7)$, (b) $(A, \lambda) = (0.08, 2.7)$, (c) $(A, \lambda) = (0.4, 0.54)$, (d) $(A, \lambda) = (1.6, 8.1)$.}
    \label{fig:packing}
\end{figure}

\begin{figure}
    \centering
    \subfigure[]{\includegraphics[width=0.7\linewidth,trim={0 0 -15pt 0}]{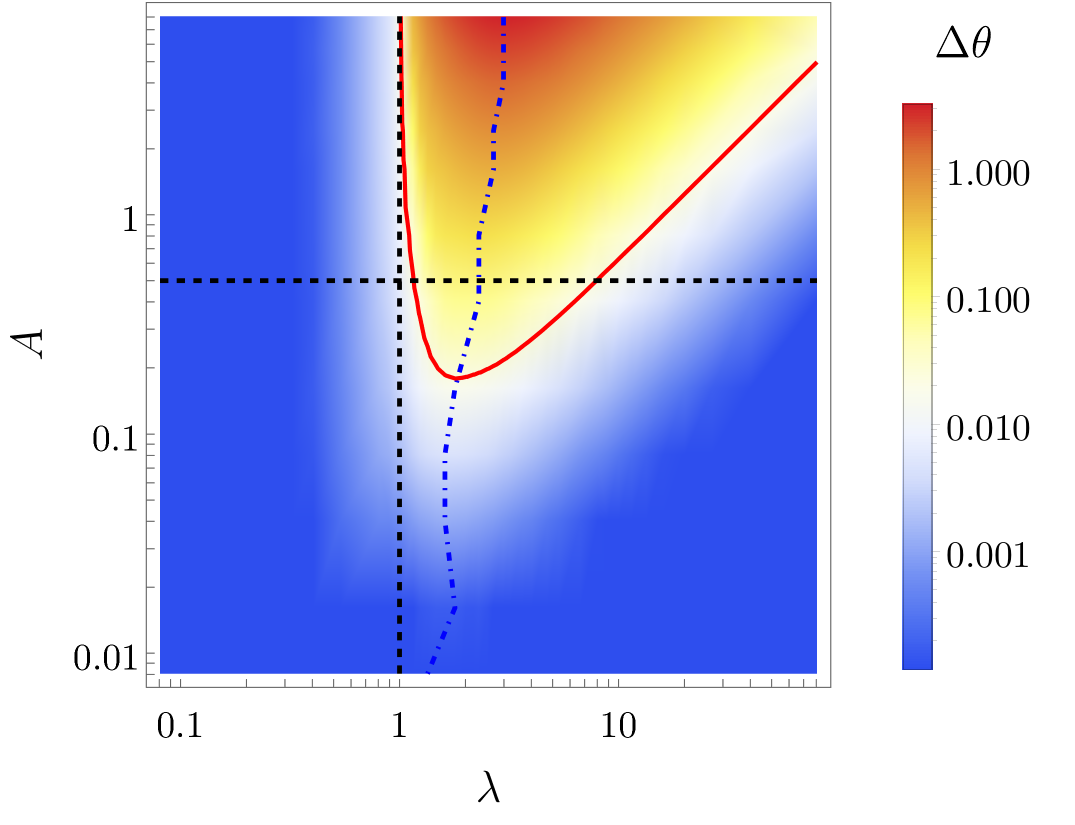}}
    \subfigure[]{ \includegraphics[width=0.7\linewidth]{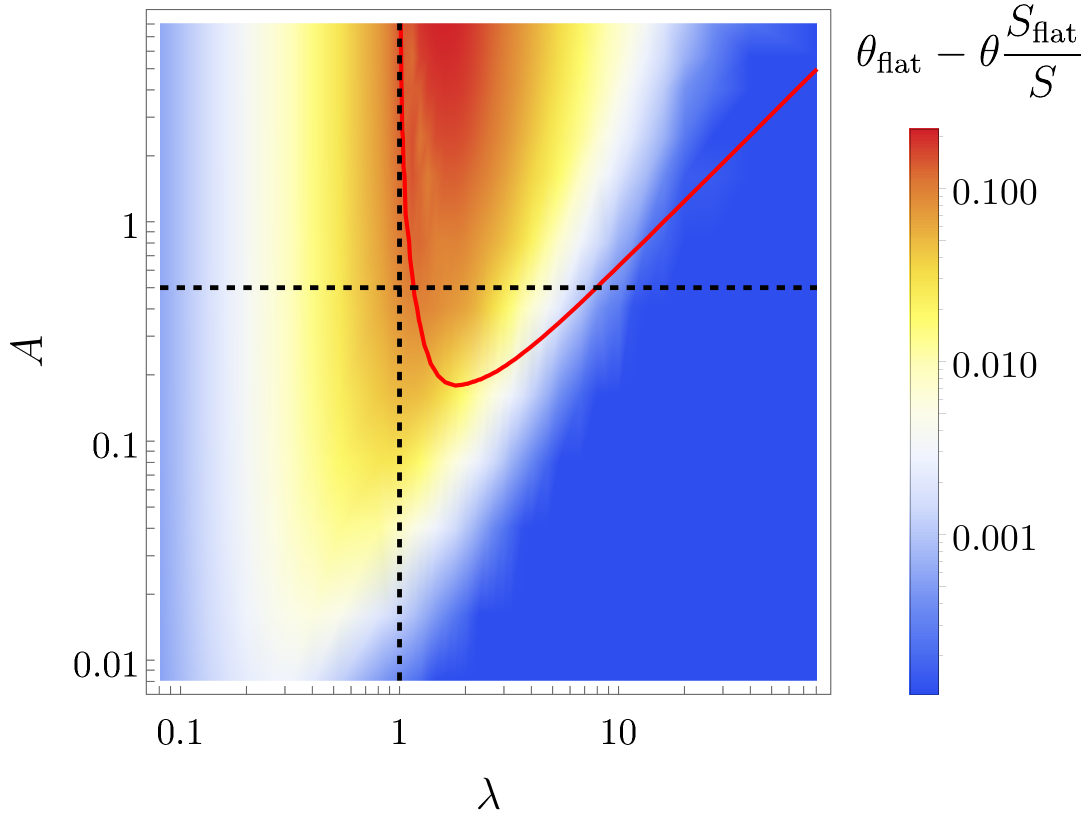}}
    
    \caption{(a) The dependence of packing fraction $\theta$ on $A$ and $\lambda$ presented as the excess above the flat surface value $\theta_\text{flat}$. The excess below $1.2 \times 10^{-4}$ is regarded as insignificant and rounded up to this value \red{and used} as a cut-off on the scale. Black dashed lines mark the values of parameters $A$ and $\lambda$ \red{yielding} surface inhomogeneities comparable with the ball size. Solid red line marks $\Delta\theta / \theta_\text{flat} \approx 4\%$ level, and blue dot-dashed line highlights $\lambda$, for which the highest $\theta$ is observed assuming a fixed value of parameter $A$. (b) The difference between the packing fraction $\theta_\text{flat}$ for a flat surface and $\theta$ for a curved surface, normalised by the area of a virtual surface $S$. Red solid line and black dashed lines correspond to the ones from panel (a).}
    \label{fig:theta}
\end{figure}

\red{Examples (fragments)} of saturated packings are shown in Fig.~\ref{fig:packing}. 
\subsection{Packing density}
The \red{effect of surface shape} should manifest itself in the deviation of the mean packing fraction from the flat surface value $\theta_\text{flat} = 0.5470690(7)$ \cite{Wang1994,Zhang2013,Ciesla2018}. As the surface waviness increases, so does the area available for the adsorbate, and thus, the packing fractions should be higher. Fig.\ref{fig:theta}(a) shows the difference $\Delta\theta = \theta(A, \lambda) - \theta_\text{flat}$ on a logarithmic scale. The statistical error was on average $\sigma(\theta) \approx 0.6 \times 10^{-4}$ and all $\Delta\theta < 2\sigma(\theta)$ \red{was regarded} as background noise. The difference $\Delta\theta$ is very prominent for large amplitudes, $A > 1$, with the packing fraction even exceeding $1$ in extreme cases. For a fixed $A$, the packing fraction reaches its maximum for some moderate $\lambda_\text{max}(A)$, with $\lambda_\text{max} \approx 3$ for $A = 8$ shifting towards $\lambda_\text{max} \approx 1.5$ for lowest amplitudes (see the dot-dashed line in Fig.\ref{fig:theta}). If both $A$ and $\lambda$ are large, the surface is \red{nearly} flat on the length scales given by particles diameter $D$. Then, $\theta$ depends only on the ratio $A/\lambda$ (c.f. the level sets of the form $A=\text{const} \cdot \lambda$). When $A$ \red{decreases}, the range of $\lambda$ with the visible difference narrows down. For $A \le 0.02$ the surface becomes indistinguishable from a flat one within statistical error. Similarly, the undulation ceases to be visible for $\lambda < 0.4$, when the surface is accessible only near \red{its peak tops} (see Fig.\ref{fig:packing}c) and becomes apparently flat regardless of how large the value of $A$ is, which results in vertical level sets in this regime. Better statistics can always increase numerical precision; however, in practice, there is a limit on how accurately the packing fraction can be measured experimentally. $\Delta\theta \approx 0.02$ ($\approx 4\%$ of $\theta_\text{flat}$) appears to be a reasonable threshold value \cite{Ocwieja2018,Yang2020}. The solid, red curve bounds this region in Fig.~\ref{fig:theta}. In \red{such} case, the amplitudes $A \in [0.2, 0.5]$ smaller than the ball radius are \red{easily} detected within experimental precision by $\Delta\theta$ in the neighbourhood of $\lambda = 1.5$. On the other hand, $\Delta\theta$ does not appear to be sufficiently sensitive to detect $\lambda < 1$. 

Although $\theta$, as defined in \eqref{eq:theta}, is the quantity measured in \red{actual} experiments, \red{the question arises whether} its value originates solely from the fact the \red{the} area $S$ of a virtual surface (cf. Fig.~\ref{fig:surface}) is larger than $S_\text{flat}$ for a flat surface. If it \red{does}, $\theta$ would be equal to $\theta_\text{flat} S/S_\text{flat}$, but \red{in fact} $\theta$ is \red{smaller, which indicates} a non-trivial role of the surface structure. Fig.~\ref{fig:theta}(b) shows the dependence of $\theta_\text{flat} - \theta S_\text{flat}/S$ on $\lambda$ and $A$. The highest deviation is visible for $\lambda$ comparable with the ball size and high \red{values of} $A$. In this regime, the virtual surface has deep valleys and a particle adsorbed near the surface minimum blocks the area on both sides of the valley. The effect is suppressed for very low $\lambda$ values when the virtual surface is \red{nearly} flat. \red{Similarly}, when $A$ and $\lambda$ are large, the surface is flat at the length scale of the particle's diameter, so rescaling according to the virtual surface area gives a good approximation of $\theta$.

\red{Direct} application of these results to the detection of roughness of a real surface below an adsorption monolayer \red{needs to incorporate the limitations and simplicity} of the RSA model. For a moderate $\lambda$ and large enough $A$ it is possible that the molecules at the neighbouring ridges block the adsorption of subsequent ones in the valley, while \red{for} the RSA protocol it \red{would be} a valid placement (for the illustration cf. the rightmost particles in Fig.~\ref{fig:surface}). Therefore, the estimation of experimental surface properties may be burdened with significant uncertainty. Nevertheless, \red{within} the most interesting region, where both $A$ and $\lambda$ are smaller or comparable to $D$, this issue is \red{of lesser prominence}.
\subsection{RSA kinetics at saturation}

For two-dimensional packings of disks on a flat surface, the difference between the saturated packing fraction and instantaneous one \red{decreases} algebraically for a large enough value of $N$, according to Feder's law \cite{Pomeau1980, Swendsen1981}
\begin{equation}
    \label{eq:fl}
    \theta - \theta(N) \propto N^{-1/d},
\end{equation}
where, in this case, $d = 2$. A similar asymptotic behaviour \red{is observed} in RSA packings of $k$-dimensional hyperspheres in a $k$-dimensional volume, with $d = k$ \cite{Torquato2006, Zhang2013}, even for fractional dimensions \cite{Ciesla2013sponge}. The algebraic scaling has also been observed for \red{majority of} two-dimensional anisotropic shapes with $d = 3$ \cite{Viot1992, Shelke2007}, which suggested that $d$ denotes the number of shape's degrees of freedom \cite{Hinrichsen1986, Ciesla2013}, and many three-dimensional anisotropic shapes with $d \in [5.5, 9]$ \cite{Ciesla2018cuboids, Ciesla2019, Kubala2019}. Parameter $d$ can be easily calculated from numerical data \red{by} performing an exponential fit to $\dv*{\theta(N)}{N}$:
\begin{equation}
    \log(\dv{\theta(N)}{N}) = \text{const} - (1/d + 1)\log N.
\end{equation}
\red{In the system investigated here}, parameter $d$ is approximately 2 for the whole range of parameters studied. Deviations do not exceed $5\%$. It is interesting, because completely different effect was observed for RSA on two-dimensional meshes \cite{Ciesla2017} where, for \red{a sparce} enough mesh, a crossover between $d=2$ and $d=1$ was observed. On the other hand, \red{the currently studied} system is three-dimensional, thus for large enough $A$ and moderate $\lambda$ a shift towards $d=3$ could be expected. To summarise, the parameter $d$, in addition to the difficulty of its accurate experimental estimation, cannot be used to detect surface waviness.  

\begin{figure}[htb]
    \centering
    \subfigure[]{\includegraphics[width=0.6\linewidth]{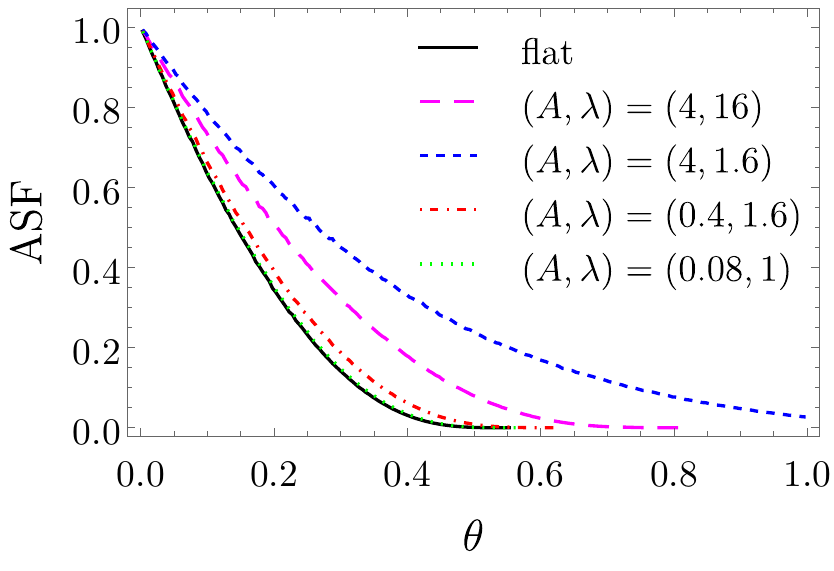}}
    \subfigure[]{\includegraphics[width=0.48\linewidth]{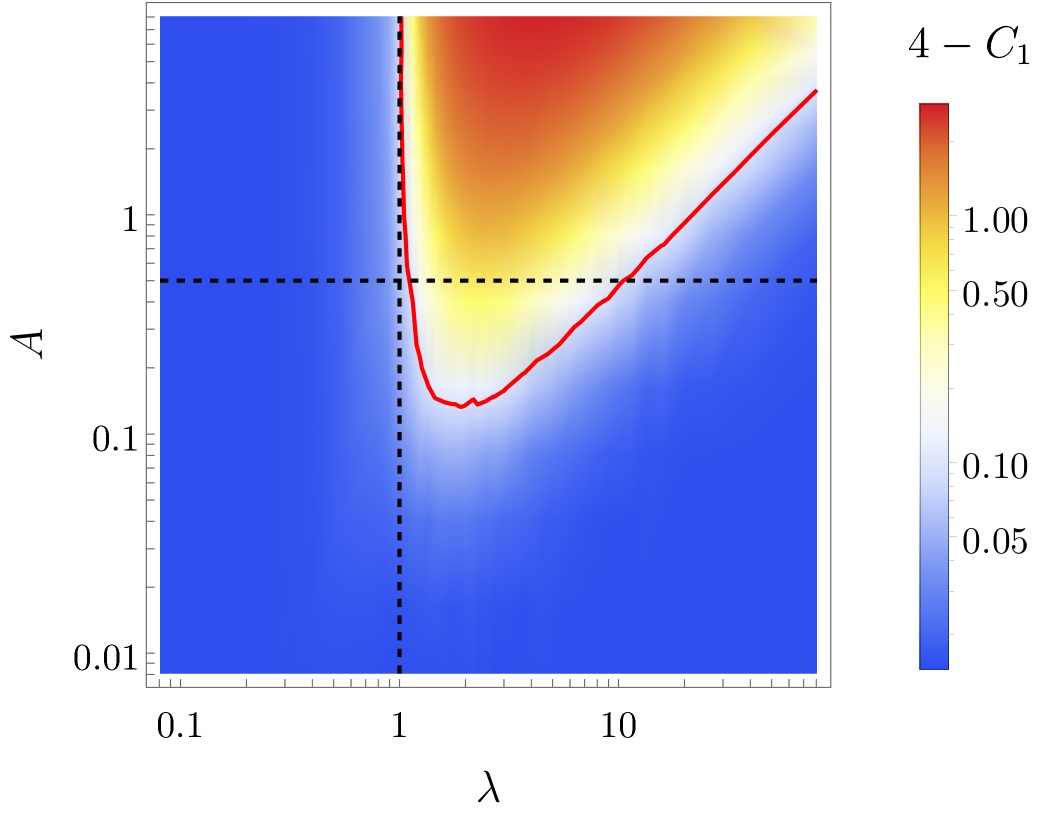}}
    \subfigure[]{\includegraphics[width=0.48\linewidth]{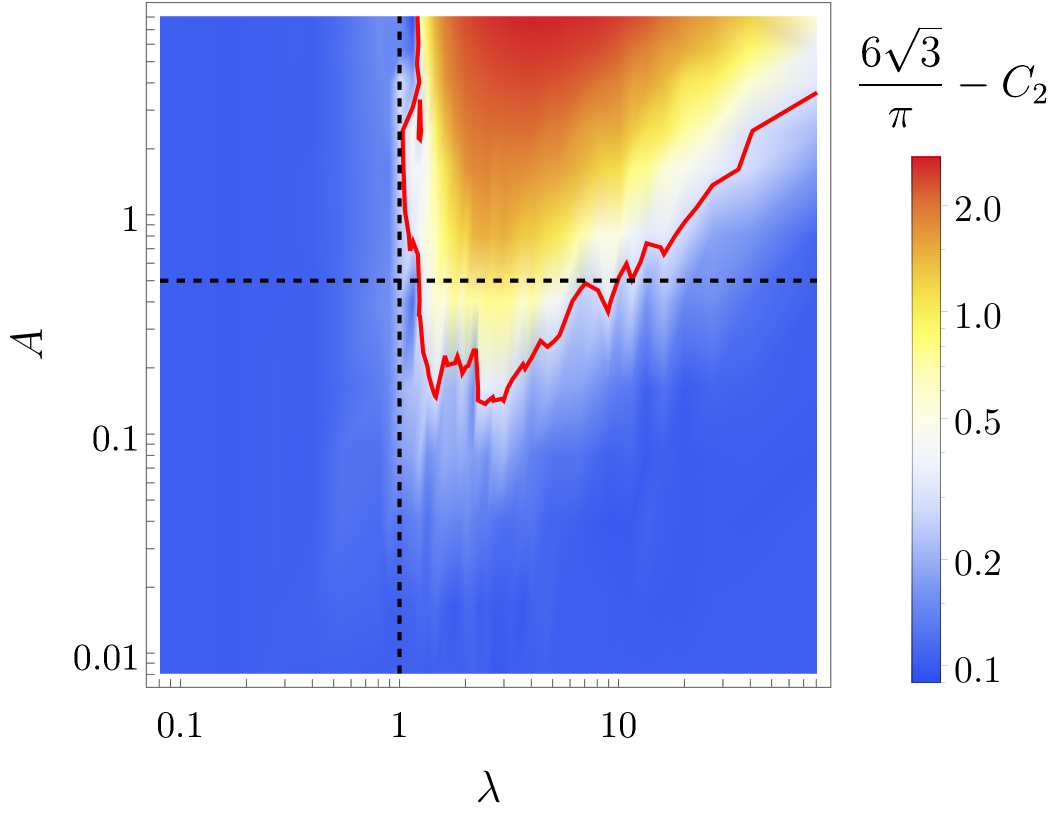}}
    \caption{The upper panel (a) shows the dependence of ASF on $\theta$ for a few selected sets of surface parameters. The lower panels show the dependence of (b) $C_1$, (c) $C_2$ on $A$ and $\lambda$. Dashed lines mark the same values of $A$ and $\lambda$ as in Fig.\ref{fig:theta}, red lines mark $\Delta C_1/C_{1,\text{flat}} \approx 5\%$ and $\Delta C_2/C_{2,\text{flat}} \approx 10\%$ levels.}
    \label{fig:ASF}
\end{figure}

\subsection{RSA kinetics at low packing density}
While Feder's law describes the kinetics of packing growth for a very large number of iterations $N$, the available surface function (ASF) is used to describe it in the limit of small $N$. ASF is defined as the probability of successful adsorption at a given $\theta$ \cite{Ricci1992,Schaaf1995}. Formally, the solution of the equation $\text{ASF}(\theta) = 0$ gives saturated packing density; however, the exact formula is known only for one-dimensional systems \cite{Wang2000}, and a perturbative approach fails to predict the correct saturated densities. Instead, \red{ASF is usually considered} up to the second order in $\theta$
\begin{equation}
    \text{ASF}(\theta) = 1 - C_1\theta + C_2\theta^2 + \mathcal{O}(\theta^3).
\end{equation}
The first two expansion coefficients are related to standard virial coefficients: $C_1 = 2B_2$, $C_2 = 2B_2^2 - \frac{3}{2}B_3$, while the higher ones require additional, non-equilibrium terms \cite{Ricci1992}. For two-dimensional disks, the analytical values of these two coefficients are $C_1 = 4$ and $C_2 = 6\sqrt{3}/\pi$ \cite{Ricci1992, Adamczyk2006}. It is interesting to check whether $C_1$ and $C_2$ values can unveil the structure of the surface. ASF can be computed from numerical simulations in the whole density range. Fig.~\ref{fig:ASF}a shows ASF for a flat surface and a few pairs of $A$, $\lambda$ parameters. In a high $\theta$ region given by $A = 4$, $\lambda = 1.6$, ASF has much higher values compared to a flat surface case in the whole density range. For $A = 0.4$, $\lambda = 1.6$, where surface variations are comparable with ball size, a visual inspection still reveals the difference. In the regime of interest, where surface variations become small, for $A = 0.08, \lambda = 1$, curved surface ASF is very similar a flat surface case. To quantify the observations, we plotted $C_1$ and $C_2$ versus $A$ and $\lambda$ in Figs.~\ref{fig:ASF}b, \ref{fig:ASF}c in a similar fashion as $\theta$ -- as deviations from flat surface values $\Delta C_1 = 4 - C_1$, $\Delta C_2 = 6\sqrt{3}/\pi - C_2$ on a logarithmic scale, using a similar two-standard-deviation cut-off. As curved surfaces are filled with particles \red{slower}, $C_1$ is expected to be lower (so is $C_2$, as all virial coefficient are expected to experience a similar qualitative scaling as $B_2 = C_1/2$). Indeed, \red{the data presented in both the panels} confirm this prediction. A region of high deviations from a flat case is similar as for $\theta$. As it has been done for $\theta$, \red{experimental resolution of} 5\% and 10\% \red{was assumed} for, respectively, $C_1$ and $C_2$, and \red{the regions of a measurable deviation were located}. The sensitivity for small surface variations is slightly higher than using $\theta$, \red{revealing} the amplitudes slightly below $A = 0.15$. \red{Additionally}, the measurement of $C_1$ and $C_2$ should be easier than \red{of} saturated packing fraction in many experimental scenarios \cite{Schaaf1995, Adamczyk1996}, so the currently described characteristics may be more practical than the former one, ie. the packing fraction.
\begin{figure}[!htb]
    \centering
    \subfigure[]{\includegraphics[width=0.6\linewidth]{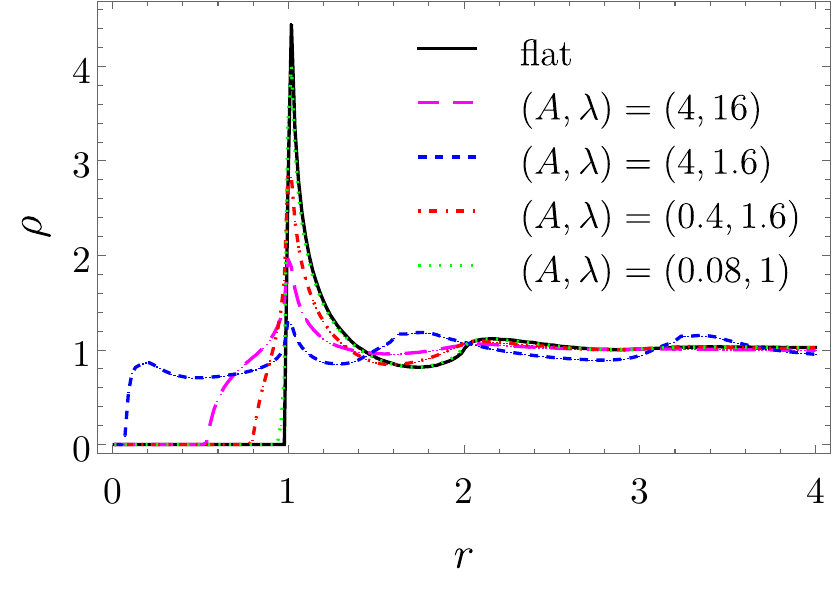}}
    \subfigure[]{\includegraphics[width=0.7\linewidth]{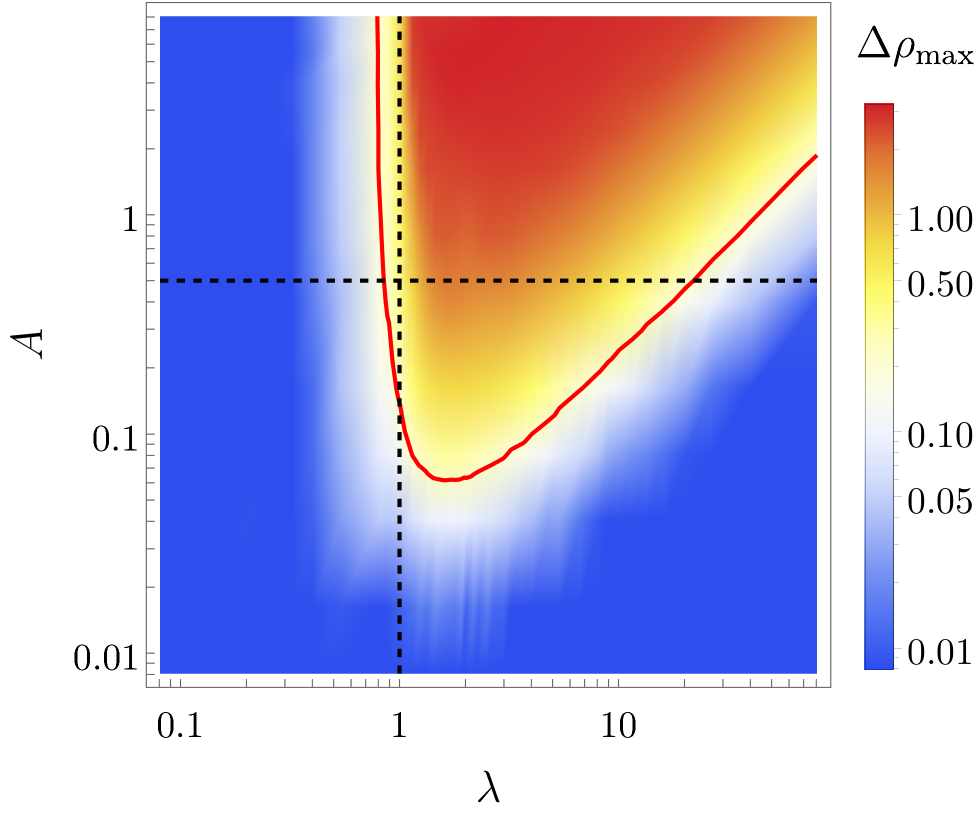}}
    \caption{The upper panel (a) shows the dependence of $\rho$ on the distance $r$ between particles for a few selected sets of surface parameters. The lower panel (b) shows the dependence of $\Delta\rho_\text{max}$ (see text) on $A$ and $\lambda$. Dashed lines mark the same values of $A$ and $\lambda$ as in Fig.\ref{fig:theta}, while solid red line marks $\Delta\rho_\text{max} / \Delta\rho_{\text{max},\text{flat}} \approx 5\%$ level.}
    \label{fig:rho}
\end{figure}
It is worth to notice that ASF is also related with Feder's law. From (\ref{eq:fl}) we get
\begin{equation}
    \frac{d\theta}{dN} \sim -N^{-\frac{1}{d}-1}.
\end{equation}
But $d\theta / dN \sim ASF(\theta(N))$ for a given packing fraction $\theta$, and, on the other hand, $N \sim [\theta(N)-\theta]^{-d}$. Thus, $ASF(\theta) \sim [\theta(N)-\theta]^{d+1}$. As shown, the same parameter $d$ governs the packing growth kinetics and the properties of ASF near saturation.
\subsection{Microstructural packing properties}
\red{The final part of the study was the analysis of} how surface inhomogeneities \red{are refleched} in correlations of particles' positions. The most common \red{measure} of translational correlations is radial distribution function, one of \red{which} equivalent definitions, convenient for numerical computations, reads as
\begin{equation}
    \rho(r) = \left\langle \frac{\dd{N}(r, r+\dd{r})}{2\pi r \dd{r}\theta} \right\rangle_{\dd{r} \to 0},
\end{equation}
where $\dd{N}(r, r+\dd{r})$ is the number of particles \red{with the} distance from each other \red{within} the range $[r, r+\dd{r}]$ and the averaging is done over all pairs of particles. Typically, in RSA packings, $\rho(r)$ \red{shows} a series of maxima and minima superexponentially \red{decreasing} to $\rho = 1$ \cite{Bonnier1994}. Moreover, in case of hypersphere packing, logarithmic divergence of $\rho(r)$ near a contact distance $D$ \red{is observed} \cite{Pomeau1980, Swendsen1981, Zhang2013}:
\begin{equation}
    \rho(r) \propto \log(\frac{r}{D} - 1).
\end{equation}
A few $\rho(r)$ curves for a selected set of parameters are \red{presented} in Fig.\ref{fig:rho}a. For a flat surface, the minimal distance giving nonzero correlations is given by touching balls and equals $r=D=1$. On a curved surface, projections of balls on $xy$ plane can significantly overlap when a local surface slope is high, resulting in a minimal distance much lower than 1, as \red{is shown} for $A = 4$ and $\lambda=1.6$. Moreover, one would expect increased correlations for $r$ being integer multiples of a sine wavelength $\lambda$. Indeed, for a high amplitude $A = 4$ and $\lambda=1.6$ maxima \red{are observed} near $r = k\lambda = 1.6, 3.2, \dots$. However, for $A = 0.4$ those maxima are \red{no longer} visible and the main difference compared to a flat surface case is the shape of the first maximum. For $A = 0.08$, $\rho(r)$ dependence almost completely overlaps with the curve for flat surface, apart from the overall maximal value. This maximal value $\rho_\text{max}$ \red{appears} to be the best indicator of surface inhomogeneity. First, \red{it is necessary} to observe that if even a slight anisotropy \red{is introduced} to the packing, the weak, logarithmic divergence \red{will regularise}. \red{It is due to the fact} that $\rho(r)$ near minimal distance is a superposition of many logarithmically divergent $\rho_D(r) \propto \log(r/D - 1)$ functions with different $D$ originating from the dependence of a minimal distance on particles' orientations. As a $0^\text{th}$ order approximation, \red{it can be assumed} that all $\rho_D(r)$ have the same weight, which gives
\begin{align}
    \rho(r) &\propto \int_{D_\text{min}}^{r} \dd{D} \log(r/D - 1) \\
    &= (r - D_\text{min})\log(1 - \frac{D_\text{min}}{r}) + D_\text{min} \log(\frac{D_\text{min}}{r}). \notag
\end{align}
This function is finite for all $r \ge D_\text{min}$. The same reasoning \red{may} be applied to curved surfaces. \red{Based on} this observation, \red{the plot} of $\Delta\rho_\text{max} = \rho_{\text{max},\text{flat}} - \rho_\text{max}$ against $A$ and $\lambda$ \red{was prepared} -- see Fig.\ref{fig:rho}b. $\rho_{\text{max},\text{flat}}$ is mathematically ill-defined because of the divergence, but the value obtained in simulation is finite due to a finite bin size $\dd{r}$. Assuming 5\% experimental resolution, \red{it was noted} that this characteristic detects amplitudes even down to $A = 0.07$ and for the first time, \red{it was possible} to observe the wavelengths below the ball diameter, namely for $\lambda$ slightly below 1. It is important to \red{mention} that $\dd{r}$ is \red{another} parameter having potentially a significant impact on $\rho_\text{max}$ values. However, as \red{it was checked}, the actual quantity of interest, namely the difference $\Delta\rho_\text{max}$ is not affected as much, especially in the region of interest, i.e. for small surface variations, near the assumed experimental resolution. \red{However, it has to be kept} in mind that the range of $\lambda$ and $A$ with visible inhomogeneity is \red{more distinct} when $\dd{r}$ is decreased, so it is vital to measure the positions of particles accurately. As the \red{centres of particles} can be \red{tracked with greater precision} than their size, $\dd{r}$ \red{was assumed to} equal 5\% of the ball's diameter.

\begin{figure}[htbp]
    \centering
    \subfigure[]{\includegraphics[width=0.65\linewidth]{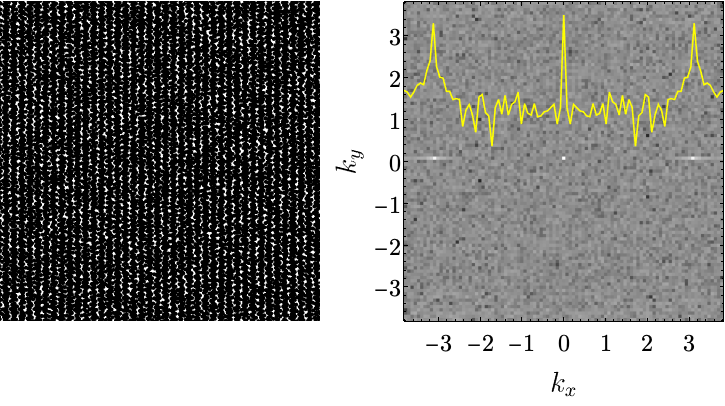}}
    \subfigure[]{\includegraphics[width=0.65\linewidth]{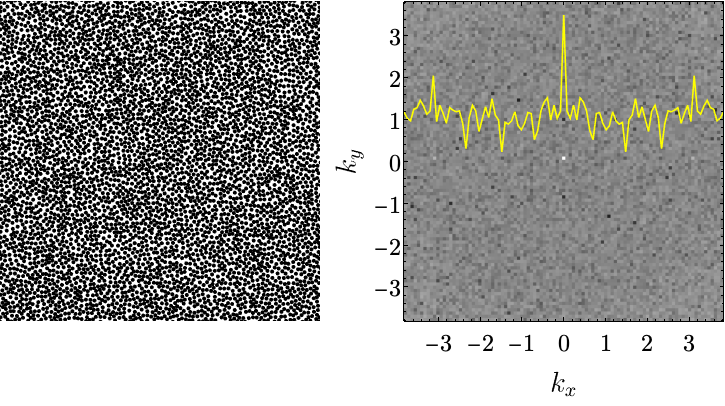}}
    \subfigure[]{\includegraphics[width=0.65\linewidth]{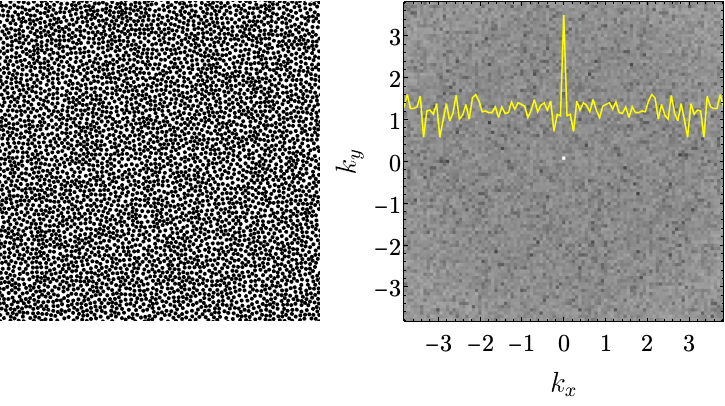}}
    \caption{Fourier transforms of orthogonal projections of a packing onto $xy$ plane for $\lambda = 2$ and various amplitudes (a) $A = 4$, (b) $A = 0.4$, (c) $A = 0.08$. Left panels depict the original packing images, while right panels -- Fourier counterparts. The brightness corresponds to the logarithm of the amplitude $W$ and the yellow solid line is the plot of $W_{k_y=0}(k_x)$ dependence with an arbitrary $y$ axis scale.}
    \label{fig:FFT}
\end{figure}

As the direct measurements of density autocorrelation function in experiments can be inaccurate and laborious, inhomogeneities may be identified using two dimensional Fast Fourier Transform of the image containing obtained adsorption monolayer \cite{Ciesla2017}. The Fourier images, together with original real space source images are presented in Fig.\ref{fig:FFT} for various amplitudes $A$ and $\lambda=2$. The axes are scaled to represent Fourier mode wave vectors $k_x$, $k_y$. As there is no modulation in $y$ direction, the `signal' -- bright spots on the image -- emerges from a noisy background only for $k_y=0$. All Fourier images contain a peak in $k_x = k_y = 0$ \red{of height depending} on the average brightness of the source image. For a high amplitude $A=4$, the periodicity in the packing is clearly visible on the real space image and \red{a well pronounced additional peak is observed} for $k_x \approx \pm 3.1$, \red{revealing} the surface wavelength $\lambda = 2$. For $A=0.4$ the real space image appears \red{to be} uniform, but FFT again reveals peaks near $k_x \approx \pm 3.1$, however, they are significantly weaker. For $A=0.08$ no signal apart from zero mode is observed. Unfortunately, FFT images of packings do not reveal surface inhomogeneities on a scale smaller than the particle size. Nevertheless, surface structure with characteristic length comparable with particle diameter is visible in FFT images of packings, which are directly accessible in the experiment.

\section{Summary}
\red{The research team} have developed the algorithm to generate saturated RSA packings of spheres on wavy surfaces, and then used it to study the properties of such packings for various amplitudes and periods of waviness. \red{It was measured} how packing fraction, the kinetics of packing growth, and microstructural properties described by the two-point density correlation function depend on these parameters. The deviations from the values reported for the flat surface grow with the waviness amplitude. For wavelengths much larger than the amplitude, the surface is indistinguishable from the flat one. Similarly, for short periods comparable to sphere diameter, the spheres are placed only close to the \red{tops of surface peaks} as they are unable to penetrate valleys between them rendering the surface apparently flat. Interestingly, the only property that \red{weakly} depends on surface waviness is the kinetics of packing growth near the saturation limit, where for the whole range of studied parameters it is governed by the exponent $-1/2$, typical for RSA of disks on a two-dimensional surface. Since RSA packings often effectively model monolayers created in irreversible adsorption experiments, the results may be used in some specific cases to detect a non-flat surface below the monolayer.        
\section*{Acknowledgements}
This work was supported by grant no.\ 0108/DIA/2020/49 of Ministry of Science and Higher Education, Poland and by grant no.\ 2016/23/B/ST3/01145 of the National Science Center, Poland. Numerical simulations were carried out with the support of the Interdisciplinary Center for Mathematical and Computational Modeling (ICM) at the University of Warsaw under grant no.\ GB76-1.

\bibliographystyle{iopart-num}
\bibliography{main}

\end{document}